# EVALUATING THE PREDICTED RELIABILITY OF MECHATRONIC SYSTEMS: STATE OF THE ART

N. Bensaid Amrani[1], L. Saintis[2], D. Sarsri[3], and M. Barreau[4].

[1] National school of applied sciences, ENSA-Tangier, BP 1818 Tangier, Morocco.
nabil.bensaidamrani@etud.univ-angers.fr.

[2] Angevin Laboratory for Research in Systems Engineering, LARIS, University of Angers, France.
laurent.saintis@univ-angers.fr

[3] National school of applied sciences, ENSA-Tangier BP 1818 Tangier, Morocco.
dsarsri@ensat.ac.ma

[4] Angevin Laboratory for Research in Systems Engineering, LARIS, University of Angers, France.
mihaela.barreau@univ-angers.fr

## ABSTRACT

*Reliability analysis of mechatronic systems is a recent field and a dynamic branch of research. It is addressed whenever there is a need for reliable, available, and safe systems. The studies of reliability must be conducted earlier during the design phase, in order to reduce costs and the number of prototypes required in the validation of the system. The process of reliability is then deployed throughout the full cycle of development. This process is broken down into three major phases: the predictive reliability, the experimental reliability and operational reliability. The main objective of this article is a kind of portrayal of the various studies enabling a noteworthy mastery of the predictive reliability. The weak points are highlighted. Presenting an overview of all the quantitative and qualitative approaches concerned with modelling and evaluating the prediction of reliability is so important for future reliability studies and for academic research to come up with new methods and tools. The mechatronic system is a hybrid system; it is dynamic, reconfigurable, and interactive. The modeling carried out of reliability prediction must take into account these criteria. Several methodologies have been developed in this track of research. In this regard, the aforementioned methodologies will be analytically sketched in this paper.*

## KEYS WORD

*Functional analysis; failure; dysfunctional analysis; mechatronic; reliability engineering; modeling qualitative analysis; stochastic model; Bayesian Network; Petri nets; Dynamic; Hybrid; Interactive; Reconfigurable; FMEA, redundancy.*

## 1. INTRODUCTION

The reliability of mechatronic systems represents an actual challenge for the future products. It is, nevertheless, still developing as a field, which is pulled up by technology and the needs of the market. The mechatronic approach is characterized mainly by the feature of coupling, between different technologies and different scientific fields of knowledge.

This inflation of the complexity at all levels increases the risk of malfunction, the methods and the tools available to the designers which enable them to master reliability, are very various and often too specific and sophisticated for a systematic use and on an industrial scale within a mechatronic framework.

The founding principle of the mechatronics manifests itself in exploiting to the maximum these couplings to offer a higher economic and technical performance; hence, sources of value being added. The increase in the levels of the couplings inevitably results in an explosion of the complexity of the systems, their control, and of the processes of design and manufacture. Before addressing "the mechatronic reliability" which represents the central theme of this paper, it would be favorable to sketch some definitions of the term "mechatronics".

French standard NF E 01-010 [1] defines "mechatronics" as an "approach aiming at the synergistic integration of mechanics, electronics, control theory, and computer science within product design and manufacturing, in order to improve and/or optimize its functionality.

According to Elsevier [2], mechatronics is the synergistic combination of mechanical precision, engineering, electronic control and systems associated with the design of products and manufacturing processes. It relates to the design of systems, devices and products aimed at achieving an optimal balance between basic mechanical structure and its overall control. The purpose of this paper is to provide rapid publication of topical papers featuring practical developments in mechatronics. It will cover a wide range of application areas including consumer product design, instrumentation, manufacturing methods, computer integration and process and device control, and will attract a readership from across the industrial and academic research spectrum. Particular importance will be attached to aspects of innovation in mechatronics design philosophy which illustrate the benefits obtainable by an a priori integration of functionality with embedded microprocessor control.

In the field of Predicted reliability, complexity of mechatronic systems is a major challenge; in this case, several studies have been carried out to handle effectively this problematic, respecting the norm 2626.

The Mechatronics approach is modeled in the design phase by the cycle in V, the model of development according to the cycle in V sets the positions of the different phases of development, since the specification until the validation [3]:

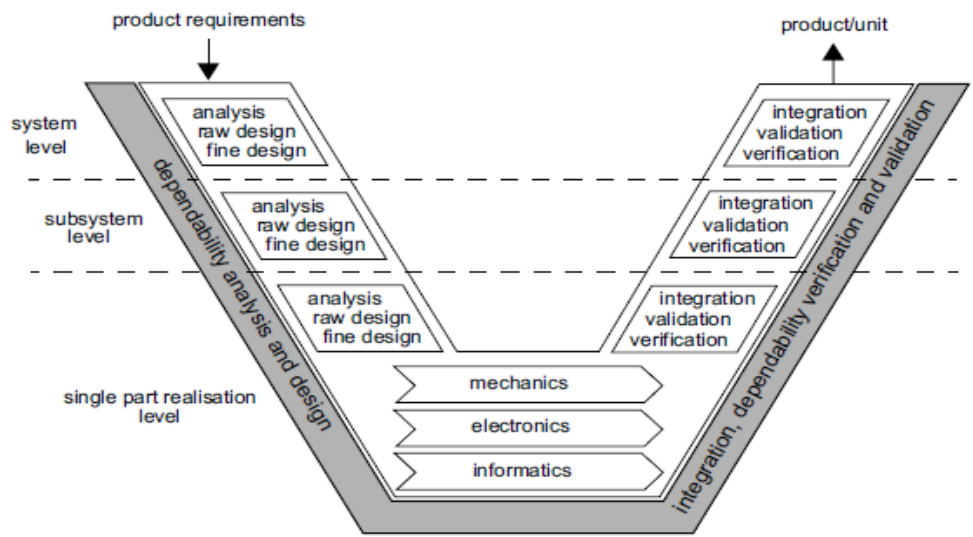

Figure 1: Integration of dependability methodology in the V-model for the development of mechatronic units.

A definition of 'reliability' is therefore necessary: it is the ability of an entity to perform the required functions in a set of given conditions during a given duration.
According to several studies [28], this concept has been treated according to the following steps of the mechatronic approach:

### 1.1 Predictive evaluation of reliability:

This phase consists of, from the beginning of the project, investigating reliability through qualitative analyses failure modes and effect analyses (FMEA,..) and quantitative (fault trees analysis (FTA), event trees (ET), integrating the different collections of data. For complex systems, it is possible to model the reliability by Petri Net. The predictive reliability allows us to take optimal orientations in the field of design.

### 1.2 Experimental evaluation of reliability:

This phase occurs as soon as the product development is sufficiently advanced and has the first prototypes. It is possible to achieve robustness tests (also called aggravated tests) in order to know the weaknesses and the margins of design. Once the product is mature (sufficient margins), a series of tests may be conducted to estimate reliability. During production, the elimination of the small deficiencies (drift process, weak component,...) is operated by screening tests.

### 1.3 Evaluation of operational reliability:

Once the product is in operation, an estimate of reliability is carried out based on the return of experience "REX" data. It is applied during the first steps and helps correct defects in design and manufacturing/production

In this article we focus this state-of-art only for the Predictive evaluation of reliability.

## 2. PROBLEMATIC:

Mechatronic systems are characterized by hybrid, dynamic, interactive and reconfigurable aspects [12]:

Hybrid systems are systems involving explicitly and simultaneously continuous phenomena and discrete events[30]
Dynamic systems are characterized by a range of functional relationships governing its constitutive components. If these relations remain frozen throughout the various tasks of the system, the system will be considered static. If, however, these relationships change during the system's functioning, the system will be regarded as a dynamic one.
The interactive nature of a system is defined by the existence of physical and/or functional interactions between the components of the system.

Finally, reconfigurable systems are systems capable of changing their internal structures to ensure the realization of the function.

What follows is an overview consisting of relatively detailed studies which are very close to the issue of the reliability of mechatronics.

## 3. EXTRACTING CRITICAL SCENARIOS FROM PETRI NETS:

The works of Sarhane Khalfaoui [4],[5] have been made in collaboration between the laboratory for analysis and Architecture of systems of CNRS and the company PSE Citroën who register in the line of work in dependability of mechatronic systems, namely establishing a methodology for estimating the reliability of these systems. In fact, Khalfaoui has developed a method of research as regards the doubted scenarios in relation to the assessment of the functional safety of mechatronic systems of the automotive realm. In his first published article [5], "Extraction of the scenarios critical from a Petri networks model using linear logic", he approached a mathematical modelling by the networks of Petri and some differential equations.

This choice was justified by the fact that this model is widely used in the modeling of discrete event systems and in studies concerned with the safe functioning of dynamic systems. From this model, a qualitative analysis is done to determine the scenarios leading to the dysfunctional event through a causality analysis underpinned by a linear logic and a backward reasoning, starting from the default state and ending with the normal operative state, in order to determine some critical scenarios, which is a consequence of the occurrence of one or a sequence of failures bringing the system into a situation of blockage in the absence of repair. The implementation of this method has been applied to two systems: the system regulating reservoirs, and an electromechanical system.

The choice of the PN Stochastic Determinist for modeling the discrete and continuous aspect of mechatronic systems is greatly adequate for the hybrid nature of these systems.

This method's results will introduce some problems, including the difficulty to define the normal state in the backward reasoning, and the impact of discrete part on the algorithm of the proposed method, in addition, the part of quantitative analysis is not treated. The disadvantages of this method lie in the multiplicity of the possible scenarios bringing about the undesirable state. Also, the problem of combinatorial explosion is avoided through extracting scenarios directly from a template PN of the system without going through the graph of accessibility and relying on graphs of proof in a linear logic allowing the management of partial orders. Otherwise, the disadvantage of this approach is when it operates solely on the discreet aspect of the template. Many inconsistent scenarios as regards the continuous dynamics are generated. Moreover, the order of occurrence of the events is not taken into account and the generated scenarios are not minimal.

On the same track, Malika Mejdouli [6] has continued the work of Khalfaoui, so she took the approach "assessment of the probability of occurrence by an approach based on linear logic and the Petri network of Predicates Transitions Differential Stochastic, In addition, she has developed a new version of the algorithm that eliminates inconsistent scenarios towards the continuous dynamics of the system. Furthermore, automation of this algorithm has been made, (this step has been proposed in the perspective generated by Khalfaoui), it has developed in JAVA a tool ESA Petri Net (Extraction & Scenarios Analyser by Petri Net model) that allows to extract critical scenarios that lead to the undesirable state from a temporal Petri net model and check some properties of the systems with calculators. Finally, this tool is applied and tested at a landing gear of an aircraft control system.

Malika Medjoudli's approach took into account the continuous aspect by temporal abstractions where necessary, and it has better characterized the scenario and its different properties as the 'minimality'. But certainly, there are other improvements to the 'minimality' of the constructed scenarios and necessarily algorithms.

# 4. THE ESTIMATION OF RELIABILITY BY QUALITATIVE AND QUANTITATIVE MODELING:

In according to A. Demri [9], [10] the estimation of reliability was made by two approaches: qualitative approach and quantitative approach. This study is based on a qualitative analysis composed of two analyses: functional and dysfunctional. The functional one allows an arborescent decomposition of the system by the methods of Structured Analysis and Design Technique S.A.D.T, or S.A in real time SART (which has the dynamic aspect lacking SADT). It can even build a correspondence of Petri Net.
Dysfunctional analysis is established by the FMEA and AEEL methods for failures, both of these analyses will build a model of PN. In an FMEA, the basic process consists of compiling lists of possible component failure modes, gathered from descriptions of each part of the system, and then trying to infer the effects of those failures on the rest of the system.

Application of laws on functional Transitions:

To obtain the time associated with functional transitions, it is necessary to represent the physical system with equations and partial derivatives taking into account the random variables. These equations allow the evolution of continuous variables of the energetic part of the system.

Application of the laws on dysfunctional Transitions:

Frequently, we associate the laws of distribution with the dysfunctional transitions following the knowledge gained on this or that component:

- For electronic exponential components.
- For software: Jelinski-Moranda model.
- For mechanical components the method mecano-reliability PHI2 to estimate the probability of failure in a temporal scale Pf (t)

As application for that proposition A.DEMRI [9] provides an application to a vehicle Antilock Brake System (ABS):

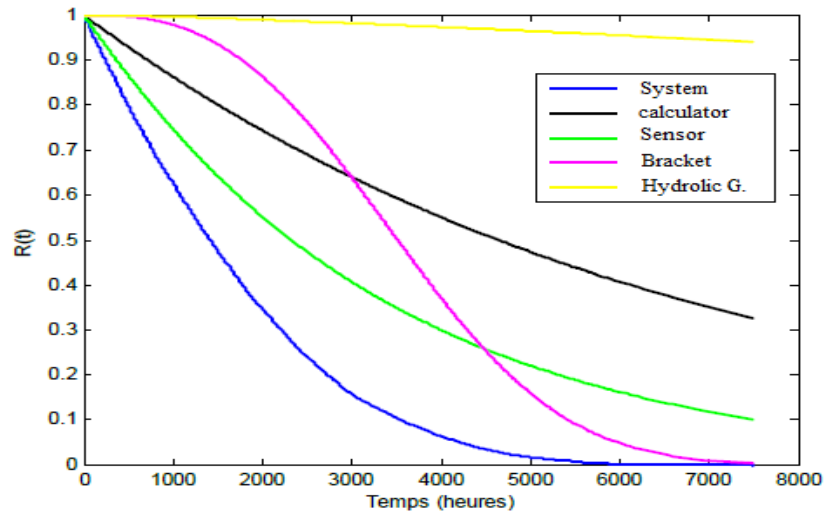

Figure 2: Reliability of the ABS system and its components.

The contribution of the A.DEMRI [9] emerged out of the need to separate the study of mechatronic reliability systems into two parts: qualitative analysis followed by a quantitative analysis. This approach has been prepared by Moncelet [21]. The main obstacle in the qualitative analysis is the combinatorial explosion of the number of the graph's states, and the quantitative analysis suffers prohibitive simulation due to taking into account by discretization of the continuous part.

## 5. THE ESTIMATION OF RELIABILITY BY QUALITATIVE AND QUANTITATIVE MODELING WITH THE RETURN OF EXPERIENCE:

Alin Gabriel [7],[8] proposed for the estimation of predictive reliability the Method Petri Net Stochastic Determined which is based on a functional modelling (providing the time of functioning); a dysfunctional modeling (giving the moments of failure). The method has been applied on the mechatronic system of the ABS. On a side note, one should have the collections of data for each component.

The collection of return of experiences 'REX' or reliability data are highly available. In practice, one often uses known databases, or even better, whenever possible, data from the manufacturers of the components. These collections are also established in the work of H.Belhadaoui.[20].

And from the databases of reliability of components, we can modelize the architecture of the system, and possibly the simulation of its operation. This method gives good results in the field of electronics but more bad results in the mechanical field, as some components do not appear in the collections of the data available.

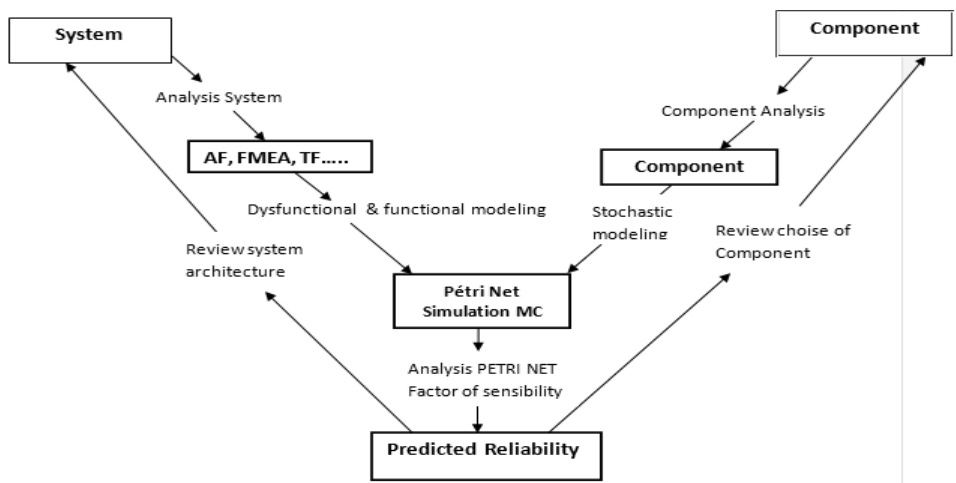

Figure 3: Evaluation Methodology predicted reliability of Mechatronic systems

This developed methodology allows to:
• model and simulate functional and dysfunctional behavior of systems
• estimate the reliability (punctual estimator and convenient interval) by simulation;
• conduct sensitivity treatment to determine the contribution of each component to the reliability of the system.

J. Gäng [25], proposes another method "HOLISTIC RELIABILITY METHOD" for mechatronic systems, it consists of a qualitative analysis, modeling (quantitative) and data collection Reliability, and other steps such as taking into account the functional failure behavior model. The problem is that one can't see the impact in quantitative modelling.
The following tables represent an example of the indexed references for electronic components [23]:

| Sources | Title | Editor |
|---|---|---|
| FIDES | Reliability methodology for electronic systems | DGA-DM/STTC/CO/477 |
| IEEE STD | IEEE Guide to the collection and presentation of Electrical. Electronic sensing component and mechanical Equipment Reliability data for nuclear Power generating stations | Institution of Electrical and Electronic engineer, New York, USA |
| MIL-HDBK-217 | Military Handbook Reliability Prediction of Electronic Equipment | United states department of defense |
| BT-HRD | Handbook for Reliability Data | British Telecommunications |
| GJB | Chinese Military Standard | Beijing ,Yuntong forever sci-Tech |

Table 1: electronic reliability Data Collection

## 6. RELIABILITY ESTIMATION BY TAKING INTO ACCOUNT THE INTERACTIONS MULTI DOMAIN:

Previous work have nevertheless ignored the other aspects of mechatronic systems, such as being reconfigurable, hybrid and interactive. These aspects are essential for the evaluation of the reliability of these systems. In this case, N.Hammouda [11] introduces the notion of organic analysis, which allows defining the architecture of the system with all the bodies and components and their interfaces to meet the expected technical functions. The main stages for the construction of organic architecture are:

- Creation of the matrix functions/components to see the functional interactions listed at the system level and at the level of each component, through the grouping into organs.
- Analysis of interactions organ/organ to determine the interactions between the various constituent organs of a primary architecture following a morphological analysis.
- Representation of organic architecture that allows visualizing the location of the various components or organs as well as their links.

The method proposed is as follows [12][13] :

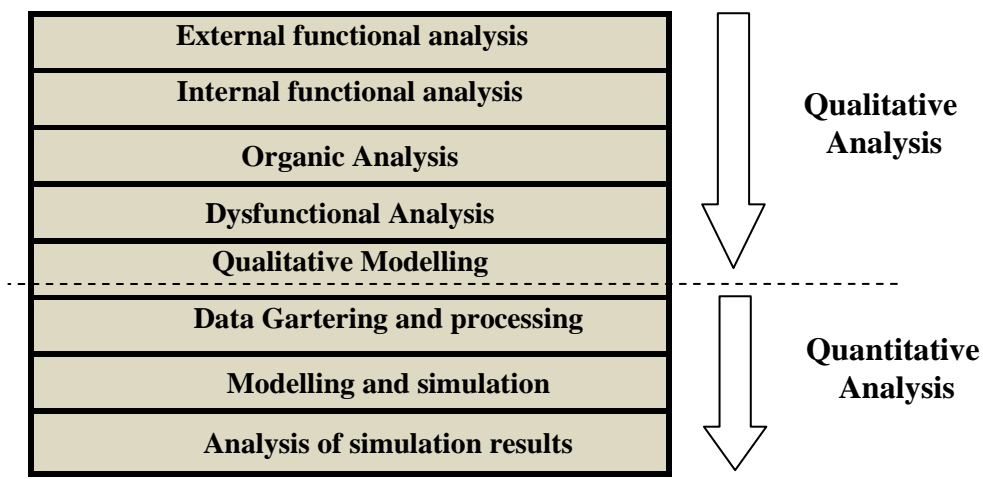

Figure 4: Estimation Approach of mechatronic reliability

At this point, the two hybrid/dynamic aspects are well covered. So for the interactive aspect N. HAMMOUDA, G. HABCHI [12], have added organic analysis functional analysis, then they injected on the PN, already made for a system (dynamic hybrid) new transitions that represent the interactions between the multi-domains.

As an application of this proposal, they choose the interaction coil/bearing as interaction between the electric domain and mechanical domain. This proposition resulted in adding value in estimating the reliability of that system "Intelligent actuator". Though, the procedure by which they have demonstrated this new tool and identified this interaction remains imprecise. This imprecision can perceived as an opportunity to conduct another study.

In the same track, J. Gäng [25] proposes a method named situation-based qualitative modeling and analysis (SQMA) representing the ex-change of information, physical quantities such as power transmission/material flow between components and human interventions as the following:

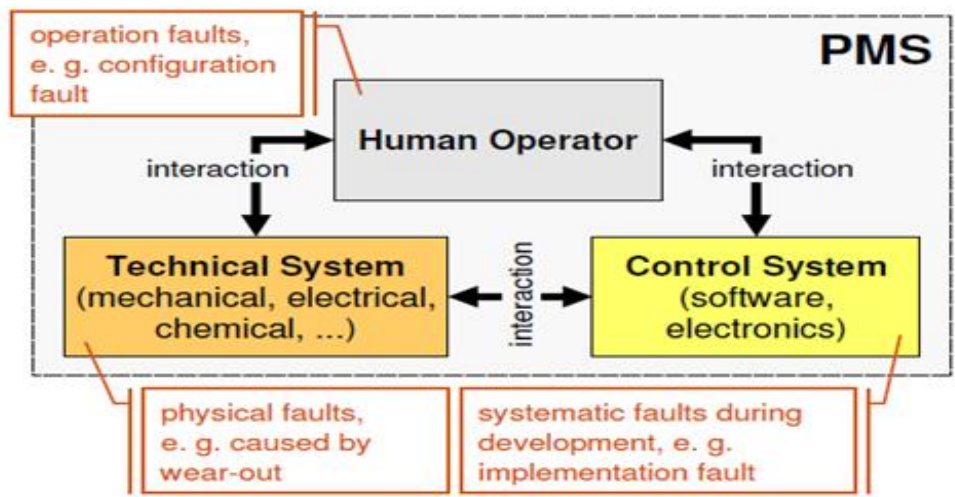

Figure 4: Structure of a computer-controlled mechatronic system

The possibility to mark the critical situations in each component is provided by SQMA tool; this method can be investigated and can be examined in tangible examples of mechatronic systems.

## 7. THE CHOICE OF TOOLS IN MODELING:

To assess and evaluate reliability, the most suitable methods for modelling and analysis of mechatronic systems are the state-transition models such as state graphs (Markov graphs, Bayesian networks) and approaches based on Petri Nets [28].

### 7.1 Statistical methods:

Statistical methods are not applicable to mechatronic systems. Fault tree does not distinguish the scenarios involving the orders of different events. [4] Indeed, a sequence of events can lead to an undesirable event while the same events. In addition, the time between two events is not explicitly taken into account in the construction of the script, so we cannot represent reconfigurations. Finally, it is not possible to take into account the temporary failures.

## 7.2 Petri Net.:

The modelling with Petri net is easier, more structured and more compact. According to A.Demri [9] the reasons that drive the choice of Petri network are:

-Modeling the different technologies of a mechatronic system;
-The use in each step of the development cycle;
-Analysis of functional and dysfunctional behavior;
-Modelling of continuous processes and discrete events
-Taking into account the dynamic aspect of the system;
-Changing their internal structures, and the specification of interactions between the various components. This tool was good applied in estimating the reliability of the photovoltaic systems in work of Remi Larond [15].

## 7.3 Bayesian Network

A Bayesian Network (BN) is a compact representation of a multivariate statistical distribution function. A BN encodes the probability density function governing a set of random variables $\{X_1…X_n\}$ by specifying a set of conditional independence statements together with a set of conditional probability functions. More specifically, a BN consists of a qualitative part, a directed acyclic graph where the nodes mirror the random variables $X_i$, and a quantitative part, the set of conditional probability functions. The advantages of this method are presented in [14], An example of a BN over the variables $\{X_1, . . . , X_5\}$ is shown in Figure 5:

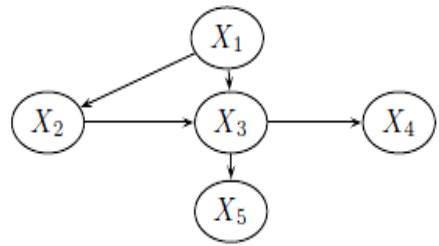

Figure 5: Example of Bayesian Networks

The underlying assumptions of conditional independence encoded in the graph allow us to calculate the joint probability function as:

$$f(x_1,\ldots,x_n) = \prod_{i=1}^{n} f(x_i|\text{pa}(x_i)).$$

## 7.4 Stochastic Hybrid –Automate

The modelling conducted by ASH allows for an evaluation of the safety of a complex system's operation, similar to the test-case considered. In addition to accessing the probability of undesirable events, it also allows for a comprehensive analysis of the emerging sequences and their respective probabilities.[24],[26].

## 8. RELIABILITY ESTIMATION TAKING INTO ACCOUNT THE IMPACT OF HUMAN FACTOR:

In the system of reliability and safety assessment, the focuses are not only the risks caused by hardware or software, but also the risks caused by ''human error''. According to Rajiv Kumar Sharma [18], a number of methods for reliability assessment such as FTA, FMEA, Petri nets, Markov analysis have been developed to model and estimate the system of reliability using the data of components. The non-probabilistic/inexact reasoning methods study problems which are not probabilistic but cause uncertainty due to the imprecision associated with the complexity of the systems as well as the vagueness of human judgment. Indeed, this uncertainty is common in a mechatronic system and none of the previous research has addressed such type of uncertainties in mechatronic systems. These methods are still developing and often use fuzzy sets, possibility theory, and belief functions.

## 9. RELIABILITY ESTIMATION WITH IDENTIFYING FAULT PROPAGATION:

Sierla et al.[17] introduced a risk-analysis methodology that can be applied at the early design phase, whose purpose is to identify fault propagation paths that cross disciplinary boundaries, and determine the combined impact of several faults in software-based automation subsystems, electric subsystems, and mechanical subsystems. Typically, error propagation analysis of hardware is based on one of the classical reliability evaluation techniques: FMEA, hazard and operability studies (HAZOP),FTA, and ET [16].

Further, they introduced a probabilistic approach to error-propagation analysis, based on Markovian representation of control flow and data flow by using UML methods. Sierla and Bryan [17] extended the work and transformed Functional Failure Identification and Propagation (FFIP) approach to safety analysis of a product line.

## 10. DISCUSSION AND SUMMARY:

Broadly speaking, the approaches proposed for estimating the predicted reliability Dynamic Hybrid have mastered aspects of a mechatronic system. The modelling of these aspects either by RDP or BN remains debatable, depending on the characteristics of the system to study, because there are various factors to take into consideration in this choice [14]. Of course, there must be a constant consideration of a modal choice.

First, extracting critical scenarios from Petri net model, especially by stochastic PN is well suited to the hybrid nature of these systems. Though, inconvenience is always the large number of undesirable scenarios generated. That problem is solved with the algorithm proposed by M. Malika [6] based also on linear logic and Petri Nets Predicate Transitions Differential Stochastic.

On the other hand, a large portion of the work done by A. Demri Amel, Alin Gabriel Mihalache, N.Hammouda, Barreau, M.,[22] is based on a qualitative approach, including functional and dysfunctional analysis. Modeling with RDP represents behavioral mode of the system, within which it is necessary to associate time with different transitions in this network. It also involves a law of probability of failure for dysfunctional transitions, and a quantitative approach based on the reliability diagram method. The difference between these works lies in choosing the efficacy kind of PN, maybe PN Stochastic, or other…

The act of introducing the experiential data as basic parameters materializes the estimation of mechatronic systems, upon which we lay emphasis for the relevance of its data. Integrating the sensitivity factor is also used to observe the influence of the input data on the reliability and even judge the influence of variables of different models.

For the interactive and reconfigurable aspects, few studies addressed mechatronics' reliability with the consideration of these two aspects. N. Hammouda had added the organic analysis to the functional one. He also introduced a new proposal for the integration of new transitions in a dysfunctional model RDP, and then repeating the calculation of the system's reliability. This approach yielded relevant results; nevertheless, it is unclear to in infer the scientific basis upon which the identification of these interactions was built.

These interactions are classified into two types [14]:

- Physical interactions (pressurization, heating, heat exchange...)
- Interactions Software / hardware, these interactions are due to the presence of communication networks, multitasking type of treatment, multiplexing in the command.

Finally, the overview of the existing approaches to fault propagation analysis has revealed that there are several fault propagation models that can be theoretically adapted to the analysis of mechatronic systems

## 11. CONCLUSION AND PERSPECTIVES:

In this article we have dealt with some methods and approaches for assessing the reliability of systems. These have produced results to better understand the hybrid, dynamic, interactive and reconfigurable aspects of mechatronic systems. However, we have noticed the insufficiency of results in some aspects, leaving the way open for a possible extension of our research. As a way to conclude, we wish to clarify that through the studies presented:

The estimation of systems' reliability is a valuable tool which helps to better define a maintenance strategy.

Given the complexity of the system due to the integration of different technologies (mechanical, electrical), we can also work on other factors including:
- Climate: moisture.
- Temperature
- Electricity: ON-OFF cycles, voltage, current, etc.
- Mechanics: torsion/flexion, chocs mécaniques, vibrations, etc.
- Another factor specific to products, as is the case with electronic products, there can be found electrostatic discharges, magnetic/electrical fields, radiation.

In general, the conception of mechatronic systems with extensive integration of different technologies requires from the beginning of the study the integration of different domains (mechanical, electrical, software, automatic) and results in a complex process, which affects the reliability of this system's conception

So, in the same way, we propose another new tool to evaluate these interactions in our other article, also for the interactions Software / hardware in future work.

As regards the reconfigurable aspect, the estimation of reliability in reconfigurable systems is a line of research still in its early stages. On our part, we have conducted a research on the same theme, but to no avail; especially for a reconfigurable mechatronic system

Indeed, the collection of data return of experience in mechatronic field still a major challenge for the research, for this field is still in its early stages of development, so we propose also to create a new pertinent database for mechatronic's components.

Finally, taking into account the uncertainty implied in evaluating the predicted reliability is so important for the development of the tools and methods in this field.

**Authors**

BEN SAID AMRANI NABIL,

Phd Student in Reliability of Mechatronic Systems in collaboration between:
The Angevin Laboratory for Research in Systems Engineering, LARIS,
University of Angers, France and the laboratory of innovative technology
In National school of applied sciences, ENSA-Tangier,Morocco.
Engineer in mechatronics, National school of applied sciences of Tetouan
Professor of engineering science.

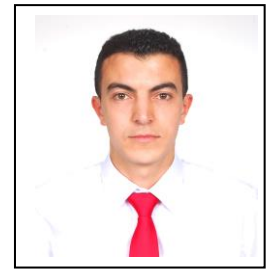